\documentclass[aip,preprint, 10pt]{revtex4}
\usepackage[LGR,T1]{fontenc}
\usepackage{textcomp}
\usepackage{amsmath}
\usepackage{amssymb}
\usepackage{graphicx}
\usepackage{esint}
\usepackage{sidecap}
\usepackage{epstopdf}
\usepackage[british,UKenglish,USenglish,english,american]{babel}
\usepackage{color}

\usepackage[footnotesize]{caption} 

\begin{document}
\title{Electrostatics of Two-Dimensional Lateral Junctions}

\author{Ferney A. Chaves and David Jim\'enez}

\affiliation{Departament d\textquoteright{}Enginyeria Electr\`onica, Escola
d\textquoteright{}Enginyeria, Universitat Aut\`onoma de Barcelona, Campus
UAB, 08193 Bellaterra, Spain. }

\begin{abstract}
The increasing technological control of two-dimensional materials has allowed the demonstration of 2D lateral junctions exhibiting unique properties that might serve as the basis for a new generation of 2D electronic and optoelectronic devices. Notably, the chemically doped MoS$_2$ homojunction, the WSe$_2$-MoS$_2$ monolayer and  MoS$_2$ monolayer/multilayer heterojunctions, have been demonstrated. Here we report the investigation of 2D lateral junction electrostatics, which differs from the bulk case because of the weaker screening, producing a much longer transition region between the space charge region and the quasi-neutral region, making inappropriate the use of the complete-depletion region approximation. For such a purpose we have developed a method based on the conformal mapping technique to solve the 2D electrostatics, widely applicable to every kind of junctions, giving accurate results for even large asymmetric charge distribution scenarios.
\end{abstract}
\maketitle

\begin{figure}[h!]
\centering 
\includegraphics[scale=0.45]{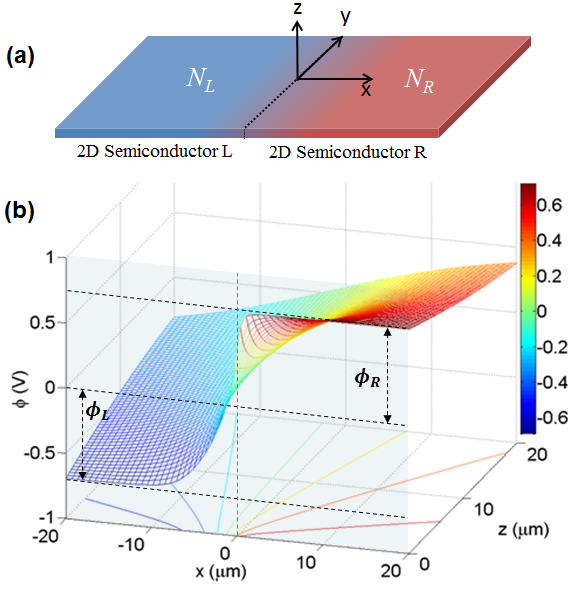}
\caption{ (a) Sketch of the 2D junction considered in this work. (b) Electrostatic potential in the $xz$-plane of an exemplary 2D lateral $pN$ homojunction. Capital letters indicate heavier doping.}
\label{figure1}
\end{figure}

\section{Introduction}
Two-dimensional (2D) layered materials like graphene and transition metal dichalcogenides, have caused a tremendous impact in the field of nanotechnology and have encouraged researchers to develop a new generation of 2D crystal based electronic devices \cite{Fiori1}. Those devices include ultrathin field-effect transistors (FETs) \cite{Radisav, Frank}, barristors \cite{Wonjae,Yang} and optoelectronic devices \cite{Choi,Lee, Bernardi}. Similar to the bulk 3D devices, these devices can contain 2D junctions as essential part of their structure. The 2D junction can be integrated into a device either by using an out-of-plane configuration featuring vertically stacked Van der Waals-bounded layers \cite{Lee} or in-plane configuration spanning adjacent regions of different composition or doping in the covalently bounded 2D plane \cite{Jariwala, Britnell, Li, Tsai}. They can either be a homojunction, as the lateral MoS$_2$ $pn$ junction \cite{Choi}, or a heterojunction, as the monolayer WSe$_2$-MoS$_2$ lateral $pn$ junction \cite{Li}. According to the majority carrier type on each side of the junction, they can be anisotype ($pN$ or $nP$) or isotype, ($pP$ or $nN$) \cite{Sun, Tosun} where the capital letter indicates stronger doping. The 2D lateral $pn$ junction is particularly useful for optoelectronics because the built-in potential, developed in both the space charge region and partially depleted transition region, readily separates and drives the photogenerated e-h pairs to generate a photocurrent. In order to understand the properties of 2D crystal based electronic devices an accurate investigation of the electrostatics of 2D lateral junctions is needed. 

Similar to bulk (3D) junctions, when a 2D lateral junction, consisting of two similar or different 2D doped semiconducting materials is formed (see Fig. \ref{figure1}a), part of the majority carriers move to the adjacent region until an equilibrium state is reached. Thus, in thermal equilibrium, the condition of zero net electron and hole current requires that the Fermi level must be constant throughout the sample. Owing to this fact, an in-plane electric field arises, the energy bands are bent and surface-charge depleted layers are formed near the interface. However, some interesting differences appear in the 2D lateral junction compared with the 3D case. In the 3D $pn$ junction a strong screening of charge carriers does exist resulting in a very narrow transition region between the space-charge region and the quasi-neutral region, so the transition region width can be disregarded and complete-depletion approximation justified. However, as for the 2D lateral junction, a significant out-of-plane electric field arises, leading to weaker screening of charge carriers, which results in much longer transition regions. In this scenario of partial depletion in a significant part of the device, is not physically correct to make the depletion region approximation \cite{Yu}. 

The impact of the dimensionality in the electrostatics of homojunctions has been investigated in Ref. \cite{Ilati}. There in, in addition to the 3D and 2D cases, the 1D case has been considered, which is relevant for devices based on carbon nanotubes, III-V nanowires or single molecules \cite{Franklin, Salmani, Brooke}. The main conclusion is that reduction of the dimensionality leads to a significant increase of the depletion width $W_D$, where the square root dependence of $W_D$ on the ratio $\epsilon/N$ in 3D changes to linear and exponential dependence for 2D and 1D, respectively, being $\epsilon$ the effective dielectric constant and $N$ the doping density.

A number of works have contributed to the quantitative understanding of the electrostatics in 2D lateral $pn$ junctions. Some of them have used the depletion approximation getting acceptable profiles of the electric field and electrostatic potential at expense of a bad description of the charge distribution along the transition region. For instance, Achoyan \textit{et al.} \cite{Achoyan} used a conformal mapping method to solve analytically the 2D Poisson equation for symmetric homojunctions. Trying to relax the depletion approximation to capture the effects of the long charge distribution tails, they solved the problem only under strongly degenerate conditions at reduced temperatures, limiting its applicability. Next, Peisakhovich \textit{et al.} \cite{Peisa} faced the asymmetric homojunction electrostatics by solving numerically an integral equation for the potential, but again for the strong degenerate case. Gharekhanlou \textit{et al.} \cite{Ghare} solved the problem analytically by modeling the depletion charge as a series of infinitesimally thin lines of charge, which results in incorrect in-plane electric field and electrostatic potential profiles, giving non-zero electric field in the quasi-neutral region along with non-monotonic potential profiles within the depletion region. Ilatikhameneh \textit{et al.} \cite{Ilati} reported an improved line-charge method, limited to symmetric junctions, taking into account the screening induced by mobile carriers in the quasi-neutral regions. The induced screening was produced by only a reflection of image charges, resulting in monotonic potential, but non-zero electric fields at the depletion edges. Recently, an extension of the image charge method was proposed by Nipane \textit{et al.} \cite{Nipane} for asymmetric $pn$ homojunctions, where an infinite number of reflections were considered to satisfy the zero electric field condition at the depletion edges. However, although the model gives accurate electric field and electrostatic potential, the charge distribution in the transition regions between the space-charge region and the quasi-neutral region far away from the junction, is not properly captured. Some other works have been reported not enforcing the depletion region approximation. Yu \textit{et al.} \cite{Yu} reported a numerical solution in the real space, based on a semiclassical approach for the charge distribution, correctly capturing the long $1/x$ charge tails in the transition region. Although accurate, working in the real space might be inadequate for very asymmetric junctions. Gurugubelli \textit{et al.} \cite{Gurugubelli} developed an analytical model for both symmetric and slightly asymmetric doping in nano-films. They considered an equipotential plane perpendicular to the junction, containing the interface line, so large doping asymmetries could not be treated properly, making important errors in the calculated physical magnitudes.

Given the above mentioned state-of-the-art in modeling/simulation of 2D lateral junctions, it becomes clear that a technique able to deal with large doping asymmetries, while keeping a correct physical description of the charge distribution tails in the transition regions, is needed. The technique we propose here is based on the solution of the non-linear 2D Poisson equation in a conformal space, without relying on restrictive hypothesis. This technique works well for both homojunctions and heterojunctions, encompassing both the isotype and anisotype cases.

\section{Methods}

\begin{figure}[h!]
\centering 
\includegraphics[scale=0.6]{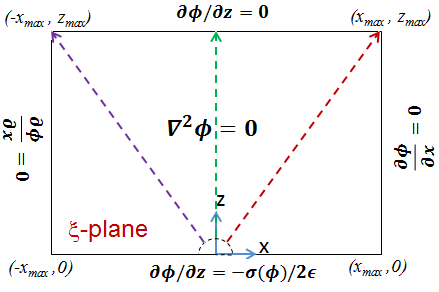}
\caption{ Computational window in the real space ($\xi$-plane) and applied boundary conditions. The dashed colored lines with constant angle are transformed in $v=$const lines in the $W$-plane (see Fig. \ref{figure3}).}
\label{figure2}
\end{figure}

\begin{figure}
\centering 
\includegraphics[scale=0.6]{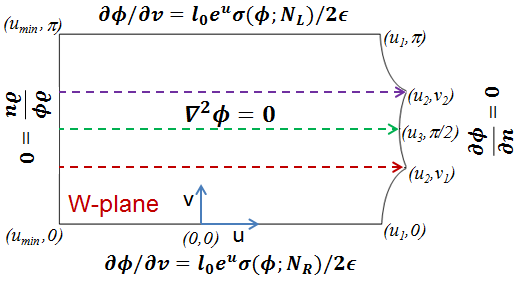}
\caption{ Computational window in the conformal space ($W$-plane) and applied boundary conditions. The dashed colored lines with $v=$const correspond to the dashed colored lines in the $\xi$-plane with constant angles (see Fig. \ref{figure2})}
\label{figure3}
\end{figure}

We propose the solution of the non-linear Poisson equation, described by the Eq. \ref{eq:poisson}, in the $xz$-plane, i.e. a plane perpendicular to the semiconducting plane, as shown in Fig. \ref{figure1}(a). Translational symmetry along $y$ and abrupt junction with uniform doping on each side are assumed. $\phi(x,z)$ represents the 2D electrostatic potential distribution (see Fig. \ref{figure1}b), $\sigma$ is the space charge density, $\epsilon$ is the effective dielectric constant of the insulating media surrounding the junction and $\delta(z)$ is the Delta function centered at $z=0$ (semiconducting plane).

\begin{equation}
-\nabla^2 \phi(x,z)=\dfrac{\sigma(\phi)}{\epsilon}\delta(z)
\label{eq:poisson}
\end{equation}

By integrating Eq. \ref{eq:poisson} along $z$ and using the Delta function definition, the problem can be transformed into the Laplace's equation:

\begin{equation}
\nabla^2 \phi(x,z)=0,
\label{eq:laplace}
\end{equation}
 
along with the boundary conditions (BCs): $\partial\phi/\partial z=-\sigma(\phi)/(2\epsilon)$ at $z=0$, and Neumann-like conditions for the rest of the borders, as shown in Fig. \ref{figure2}. The $\lambda$-side of the junction, where $\lambda$ could be left (L) or right (R), represents a 2D semiconducting material with surface doping density $N_\lambda$ and band gap energy $E_{\lambda,g}$. The total surface charge density $\sigma_{\lambda}$ in the $\lambda$-side depends on the ionized fixed charge density $N_\lambda^i$ and the 2D mobile carrier densities $p_{\lambda}$ and $n_{\lambda}$ in the valence band (VB) and conduction band (CB), respectively:

\begin{equation}
\sigma_{\lambda}(\phi)=q\left( N_{\lambda}^i+p_{\lambda}-n_{\lambda}\right), 
\label{eq:sigma}
\end{equation}

If complete ionization is assumed, then $N_\lambda^i=\pm N_\lambda$. According to the parabolic dispersion relation of the mobile carrier density in 2D semiconductors, the band-edge density of states is given by $g_{\lambda}(E)=g_{\lambda,s}g_{\lambda,v}m_{\lambda}^*/2\pi\hbar^2$, where $\hbar$ is the reduced Planck constant, $m_{\lambda}^*$, $g_{\lambda,s}$ and $g_{\lambda,v}$ are the band-edge effective mass, spin and valley degeneracy factors of the semiconducting layer in the $\lambda$ side, respectively. $p_{\lambda}$ and $n_{\lambda}$ are accurately described as $p_{\lambda}=\int_{-\infty}^{E_{\lambda,v}}g_{\lambda}[1-f(E)]dE$ and $n_{\lambda}=\int_{E_{\lambda,c}}^\infty g_{\lambda}f(E)dE$ where $E_{\lambda,v}$ and $E_{\lambda,c}$ are the band-edge energies of the VB and CB, respectively.  By using the occupation probability described by the Fermi-Dirac distribution, the electron density in the CB is given by $n_{\lambda}=g_{\lambda}k_BTln\left\lbrace 1+exp\left[ (E_f-E_{\lambda,c})/k_BT \right]  \right\rbrace $ and the hole density in the VB is $p_{\lambda}=g_{\lambda}k_BTln\left\lbrace 1+exp\left[ -(E_f-E_{\lambda,v})/k_BT \right]  \right\rbrace $, with $k_B$ the Boltzmann constant, $T$ the absolute temperature, and $E_F$ the Fermi level. In general, electrons and holes could have different effective masses. By defining the electrostatic potential $\phi$ such that $q\phi=E_F-E_i$, where $E_i$ is the semiconductor intrinsic energy, we can write:

\begin{equation}
n_{\lambda}=g_{\lambda}k_BTln\left[1+exp \left(\dfrac{q\phi-E_{\lambda,g}/2}{k_BT} \right)  \right]  
\label{eq:nn}
\end{equation}

\begin{equation}
p_{\lambda}=g_{\lambda}k_BTln\left[1+exp \left(-\dfrac{q\phi+E_{\lambda,g}/2}{k_BT} \right)  \right].  
\label{eq:pp}
\end{equation}

From the Eqs. \ref{eq:nn} and \ref{eq:pp}, under thermal equilibrium and far away of the junction (quasi-neutral regions), the in-plane potential in the $\lambda$-side must tend asymptotically to: 

\begin{equation}
\phi_{\lambda}=\pm\lbrace k_BTln\left[ e^{N_{\lambda}/g_{\lambda}k_BT }-1\right]+E_{\lambda,g}/2\rbrace/q,
\label{eq:fiLR}
\end{equation}

as shown in Fig. \ref{figure1}b. Thus, the in-plane built-in potential $\phi_{bi}$ along the junction can be expressed as $\phi_{bi}=\vert\phi_L\vert+\vert\phi_R\vert$. 

For $pn$ junctions, an asymmetry factor $r$ can be defined as the ratio between the surface doping densities $N_L$ and $N_R$ of the junction ($r=N_L/N_R$), so $r=1$ corresponds to a symmetric junction. For slightly asymmetric 2D $pn$ junctions, depletion width $W_\lambda\approx\dfrac{4\epsilon\phi_{bi}}{\pi q N_{\lambda}}$ has been estimated for each $\lambda$-side \cite{Gurugubelli, Peisa}. So, when $r\sim1$ the depletion widths at both sides of the junction are of the same order, and the numerical problem could be solved in the real space $xz$, as shown in Fig. \ref{figure2}. On the contrary, if a large asymmetry does exist ($r<<1$ or $r>>1$) the depletion widths might vary in orders of magnitude, making the numerical solution in the real space hard to get. To overcome this difficulty, we propose to solve the numerical problem in a transformed $W$-space by means of a conformal mapping (see Fig. \ref{figure3}). It is a transformation by means of an analytic function that preserves angles between every pair of curves from the complex $\xi$-plane ($\xi=x+iz$) to the complex $W$-plane ($w=u+iv$) \cite{Ward}. In mathematical physics, a harmonic function is a twice continuously differentiable function which satisfies the 2D Laplace's equation, so the potential distribution $\phi(x,z)$ is a harmonic function of $x$ and $z$.  On the other hand, every harmonic function of $x$ and $z$ transforms into a harmonic function of $u$ and $v$ under a change of variables $x+iz=f(u+iv)$, where $f$ is an analytic function. For the conformal transformation we have found appropiate the function described by

\begin{equation}
w=ln(\xi/l_0),  
\label{eq:transf}
\end{equation}

 where $l_0$ is a real scalar. Therefore, $\phi(u,v)$ is also a harmonic function satisfying the Laplace's equation in the $W$-plane.

According to the conformal transformation, the BCs expressed in the $\xi$-plane, and represented in Fig. \ref{figure2}, satisfying that the normal derivative of $\phi(x,z)$ is zero, are transformed into BCs with zero normal derivative in the $W$-plane, as shown in Fig. \ref{figure3}. As for the BCs where the normal derivative is non zero (semiconducting plane), the conformal transformation dictates that the ratio of the directional derivative of $\phi$ in the $\xi$-plane to the directional derivative of $\phi$ in the corresponding direction in the $W$-plane is $\vert dw/d\xi\vert$. So the positive and negative $x$-axes borders of the $\xi$-plane with the BC $\partial\phi/\partial z=-\sigma/2\epsilon$ are mapped into the $v=0$ axis with BC $\partial\phi/\partial v=l_0e^u\sigma_R/2\epsilon$ and $v=\pi$ axis with BC $\partial\phi/\partial v=l_0e^u\sigma_L/2\epsilon$ in the $W$-plane, respectively.  

The formula for the depletion width given above can be used as an appropriate guess for tailoring the computational window size of the Fig. \ref{figure2} . Specifically,  we have used $x_{max}=\alpha_1W_{max}$ and $z_{max}=\alpha_2W_{max}$, where $\alpha_{1(2)}$ is a large enough positive real number and $W_{max}=max(W_L, W_R)$. 
      
In Fig. \ref{figure3}, the colored dashed lines and the coordinates ($u_1$,0), ($u_2$,$v_1$), ($u_3$,$\pi$/2), ($u_2$,$v_2$), and ($u_1$,$\pi$) of the $W$-plane come from the conformal transformation of the colored dashed lines and coordinates ($x_{max}$,0), ($x_{max}$,$z_{max}$), (0,$z_{max}$), (-$x_{max}$,$z_{max}$) and (-$x_{max}$,0) in the $\xi$-plane, respectively. The parameter $l_0$ is defined as $l_0=W_{min}=min(W_L, W_R)$ and the left boundary  $(u_{min},v)$ comes from the transformation of the small dashed semicircle represented in Fig. \ref{figure2}, which has a radius $x_{min}$, where $x_{min}=W_{min}/\alpha_3$ with $\alpha_3$ a large enough positive real number.   

The solution of Laplace's equation in the W-plane with the corresponding BCs, is obtained by using an algorithm based on a Gauss-Newton iteration scheme applied to the finite element matrix coming from a finite element mesh. Once the potential $\phi(u,v)$ is got, the electrostatic potential $\phi(x,z)$ in the $\xi$-plane could be mapped back by using Eq. \ref{eq:transf}. The result is illustrated in Fig. \ref{figure1}b, including the in-plane electrostatic potential $\phi(x)=\phi(x,0)$. It is worth noticing that $\phi(x)$ tends asymptotically to $\phi_L$ and $\phi_R$ as $x$ approaches $-\infty$ and $\infty$, respectively, even though Dirichlet type conditions have not been enforced. We have used a large computational window in order to avoid any significant dependence of the in-plane potential $\phi(x)$ and the surface charge density $\sigma$ on the computational window size. For instance, the combination $\alpha_1=25$, $\alpha_2=1.5\alpha_1$ and $\alpha_3=10^3$ usually gives good results.  

\section{Results and discussions}  
\begin{figure}[h!]
\centering 
\includegraphics[scale=0.5]{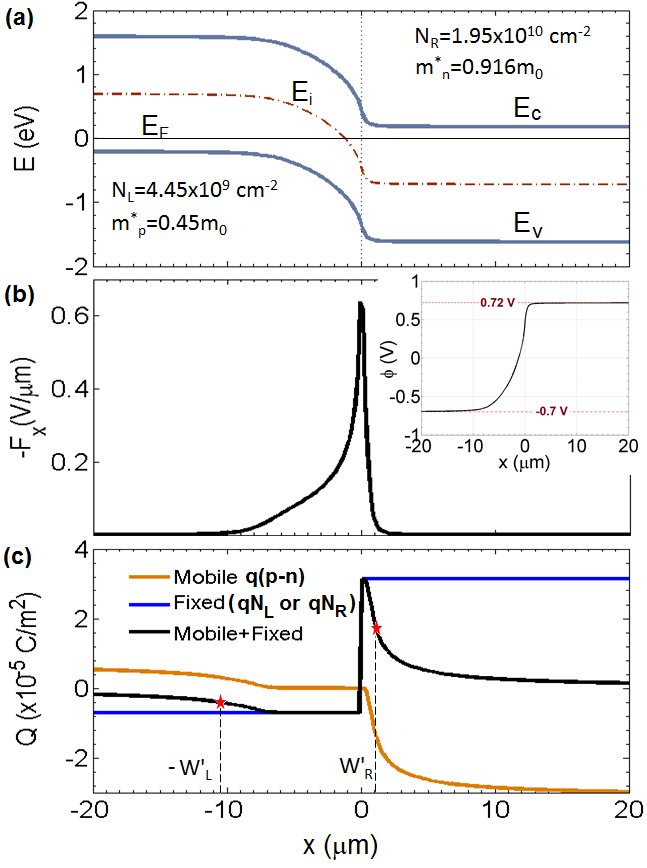}
\caption{ Electrostatics of a 2D MoS$_2$ lateral $pN$ homojunction with $r$=0.23. (a)Energy bands; (b) In-plane electric field and electrostatic potential (Inset); (c) surface density charges. $W_\lambda'$ is the effective depletion width. We have assumed a dielectric environment $\epsilon=3.9\epsilon_0$}
\label{figure4}
\end{figure}  

\begin{table}
\centering 
\resizebox{8cm}{!}{
\begin{tabular}{|c|c|c|c|c|}
  \hline 
  $E_g$ (eV) & $m_n^*$ ($m_0$) & $m_p^*$ ($m_0$) & $g_v$ & $g_s$ \\ 
  \hline 
  1.8 & 0.916 & 0.45 & 2 & 2 \\ 
  \hline 
  $N_L^i$ (cm$^{-2}$) & $N_R^i$ (cm$^{-2}$) &$r$& $\epsilon$ ($\epsilon_0$) & T (K) \\ 
  \hline 
  -4.45$\times10^9$ & 1.95$\times10^{10}$ &0.23& 3.9 & 300 \\ 
  \hline 
  \end{tabular} 
}
\caption{Parameters used to investigate the electrostatics of a 2D lateral $pN$ homojunction }
\end{table}   

\begin{figure}
\centering 
\includegraphics[scale=0.55]{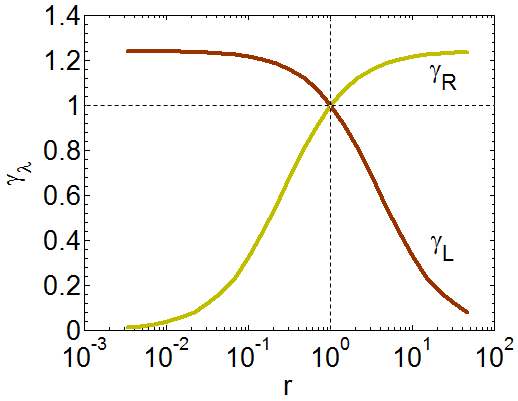}
\caption{Ratio between the effective depletion width extracted in this work ($W_\lambda'$) and the analytical prediction for the symmetric case, as a function of the asymmetry factor $r$ .}
\label{figure5}
\end{figure}

\begin{figure}
\centering 
\includegraphics[scale=0.4]{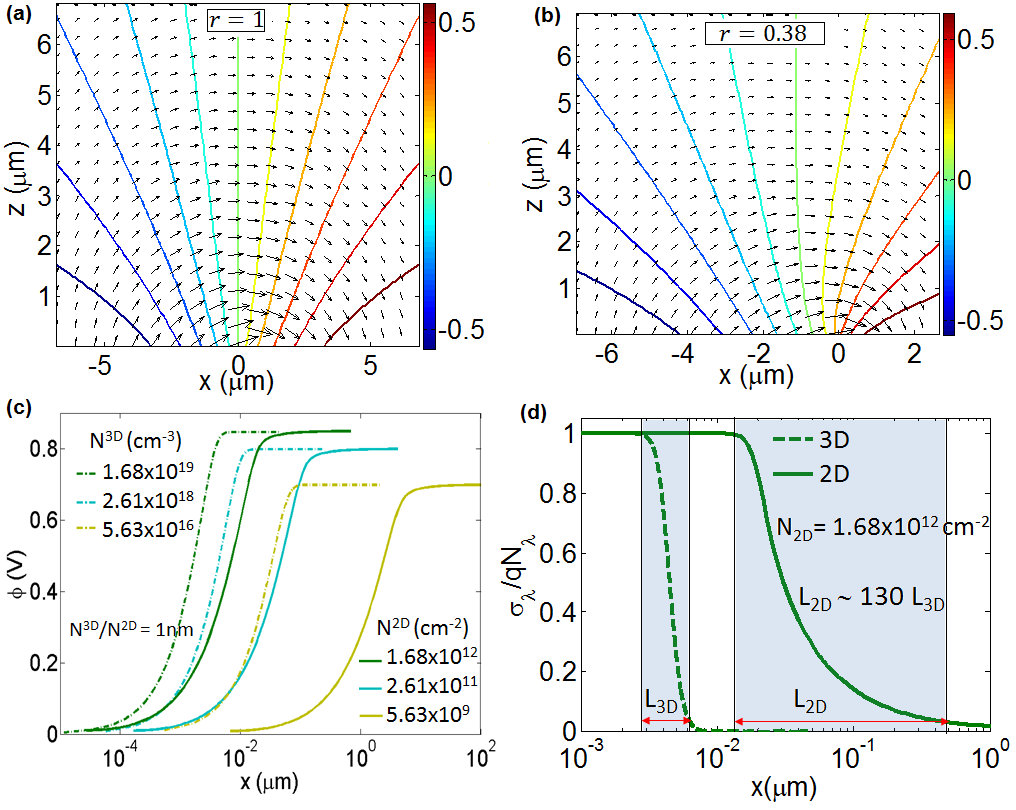}
\caption{(a)-(b) Equipotential lines and electric field of a 2D lateral $pn$ homojunction in the $xz$-plane for two values of $r$. (c) The in-plane electrostatic potential of a symmetric 2D lateral $pn$ homojunction for several surface density doping, compared with the bulk case. (d) Comparison of the transition region widths  between 2D and 3D homojunctions. Only the $n$ region is shown.}
\label{figure6}
\end{figure}

\begin{figure}
\centering 
\includegraphics[scale=0.4]{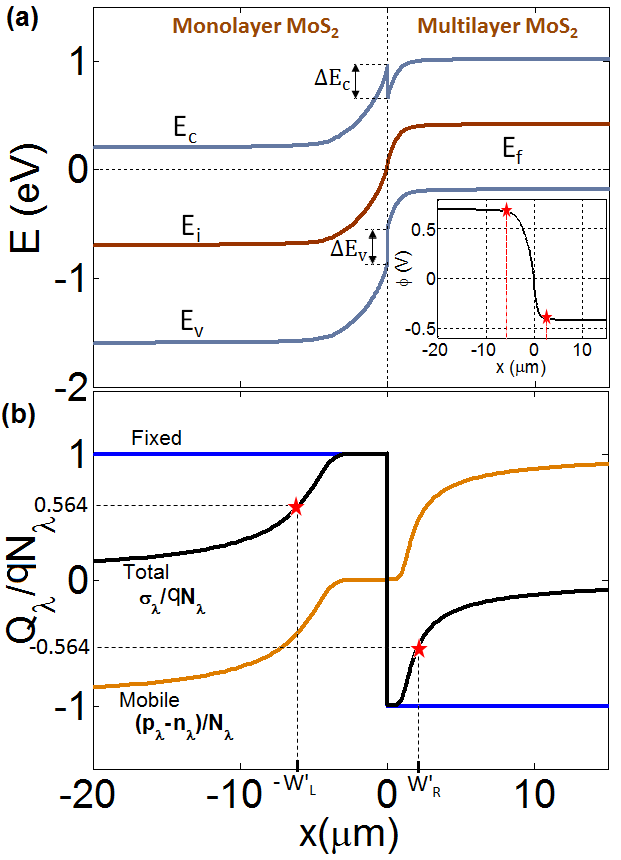}
\caption{ Electrostatics of a 2D lateral monolayer-multilayer $Np$ heterojunction. (a) Energy bands and electrostatic potential (Inset); (b) Surface charge density relative to the fixed charge at both sides of the junction. Red stars in (b) and inset of (a) indicate the relative charge and electrostatic potential at the edge of the depletion widths, respectively. }
\label{figure7}
\end{figure}

Fig. \ref{figure4} shows the electrostatics of an asymmetric 2D lateral $pN$ homojunction at room temperature, assuming a monolayer MoS$_2$ as the semiconducting material with physical parameters in the Table I. Fig. \ref{figure4}a corresponds to the profile of the minimum and maximum of the CB ($E_c$) and VB ($E_v$), respectively, whose energies are given respect to the Fermi energy $E_F$ of the junction, taken as the reference. The in-plane electric field $F_x$, electrostatic potential $\phi(x)$ and surface charge distribution are shown in the Fig. \ref{figure4}b-c for $r$=0.23. The values $0.72$V and $-0.7$V represented in the inset of Fig. \ref{figure4}b correspond to the asymptotic potential $\phi_\lambda$ given by Eq. \ref{eq:fiLR}. Given that in symmetric 2D $pn$ junctions a  partially depleted transition region exists, in which $\sigma_\lambda/qN_\lambda$ is reduced in a 43.6\% at $W_\lambda$, as shown in Fig. (S1)a of the Supplementary data (SD), we could define, using the same criterion of the 43.6\%, an effective depletion width $W_\lambda'$ for asymmetric $pn$ junctions. In terms of the in-plane potential $\phi(x)$ (see Fig. S2), for semiconducting layers with gap energy $E_{\lambda,g}$ larger than 1.2 eV, the potential at $W_\lambda$  is larger than the 93\% of its asymptotic value $\phi_\lambda$ (see Figs. (S1)b and (S2)b). Thus, $\sigma_\lambda/qN_\lambda=0.564$ could be acceptably used as a natural extension of the criterion to define $W_\lambda'$ for both homojunctions and heterojunctions. Interestingly, a general relationship between  $W_\lambda'$ and $W_\lambda$ can be expressed as $W_\lambda'=\gamma_\lambda W_\lambda$, where the dependence of $\gamma$ with $r$ has been shown in Fig. \ref{figure5} for a homojunction. Here we have assumed a fixed value of $N_L$ and variable $N_R$. In general, the stronger the asymmetry the larger the difference between $W_\lambda'$ and $W_\lambda$. The curves $\gamma_R$ and $\gamma_L$ are specular each other respect to the line given by r=1. It is worth noting that even for low asymmetries, for example $N_R=0.1N_L$ ($r=10$), $W_L'$ and $W_R'$ differ by approximately 63\% and 20\% of their predicted values $W_L$ and $W_R$, respectively. These quantitative results given by the numerical model are quite useful for benchmarking of future compact models of the 2D $pn$ junction electrostatics. Moreover, the proposed method could be helpful for designing the active region in optoelectronic devices, where maximization of photogenerated carriers or minimization of time spent by the photocarriers to reach the contacts is relevant for the specific application. In addition, the dependence of $\gamma_\lambda$ on $r$, $\phi_\lambda$ and $N_\lambda$ for several types of junctions can be seen in Figs. (S3) to (S7) of the SD. There in, we also report the dependence of $\gamma_\lambda$ with the temperature, which is very weak in the temperature range $20-400$ K. On the other hand, $\gamma_\lambda$ is sensitive to the surrounding dielectric constant so the environment could be used as a way to linearly modulate the depletion width \cite{Nipane}.

Next, by using our results, we want to highlight the above mentioned differences between the 2D and bulk junctions. The out-of-plane electrostatics is shown in the Fig. \ref{figure6}a-b for two values of $r$ in an exemplary 2D lateral homojunction, assuming $m_p=m_n=0.57m_0$. The colored lines and the tiny arrows represent the equipotential lines and the electric field across the $xz$ plane. There has been assumed translational symmetry along the $y$-axis.   Clearly the $yz$ plane at $x=0$ is an equipotential plane for the symmetric case ($r=1$) but this is not the case for the asymmetric case ($r \neq 1$). So large doping asymmetries can not be treated properly by using the previously mentioned methods which are based on the assumption of an equipotential plane at $x=0$. The behavior of $\phi(x)$ at a fixed distance $z>0$ from the semiconducting plane can be understood with the help of Fig. \ref{figure1}b. Both Fig. \ref{figure1}b and Fig. \ref{figure6}a-b show that our method allows to get the electrostatics of a $pn$ junction, not only in the plane of the semiconductors but out-of-the-plane.

Fig. \ref{figure6}c shows the in-plane electrostatic potential of a symmetric 2D lateral $pn$ homojunction compared with the 3D case, for several doping densities, and Fig. \ref{figure6}d shows the relative charge of  both cases for a fixed doping. To do a fair comparison between their depletion widths we have assumed an hypothetical bulk semiconductor made of multiple 2D layers separated a distance $d=1$ nm with a 3D carrier intrinsic concentration such that the built-it potential in both the 3D and 2D junctions is the same when $N^{3D}=N^{2D}/d$. Given that for 3D $pn$ junctions $W_\lambda\sim 1/\sqrt{N_\lambda}$ while for 2D $pn$ junctions $W_\lambda\sim 1/N_\lambda$, the difference between their depletion widths increases as the semiconductors are low doped \cite{Yu}. Clearly, the partially depleted 2D transition region ($L_{2D}$) is orders of magnitude larger than the 3D one ($L_{3D}$), so it is hard to get an accurate description of the charge distribution in 2D using the complete-depletion approximation. 

Additionaly, our proposed technique can be also applied to 2D lateral $pn$ heterojunctions. For instance, it allows us to calculate the electrostatics of a monolayer-multilayer MoS$_2$ junction \cite{Sun} with energy band gaps $E_{L,g}=1.8$ eV and $E_{R,g}=1.2$ eV for the semiconducting monolayer and multilayer, respectively. The surface doping densities assumed are $N_L=5.67\times10^{9}$ cm$^{-2}$ and $N_R=1.22\times10^{10}$ cm$^{-2}$ corresponding to $\phi_L=0.7$ V and $\phi_R=-0.42$ V, respectively, and $r$=0.46. The profiles of the CB, VB and intrinsic level energy $E_i$ are shown in the main panel of the Fig. \ref{figure7}a. For this examined heterostructure, $\Delta E_c=0.29$ eV and $\Delta E_v=-0.31$ eV, which are the discontinuities in the CB and VB, respectively, produced at the junction. Also, the inset of the Fig. \ref{figure7}a shows the resulting electrostatic potential, where the edges of the effective depletion regions have been indicated by red stars. The mobile ($q(p_\lambda-n_\lambda)$) and total surface charge relative to the fixed charge $qN_\lambda$ is shown in Fig. \ref{figure7}b along with the effective depletion width $W_\lambda'$ which are 6.15 $\mu m$ and 1.97 $\mu m$ for the monolayer and the multilayer MoS$_2$, respectively. The corresponding $\gamma_L$ and $\gamma_R$ for this heterojunction are 1.12 and 0.88, respectively. The dependence of $\gamma_\lambda$ on $N_R$ can be seen in Fig. S5 of the SD.  The edges of the effective depletion regions lie inside the transition regions, which display long $1/x$ charge tails. Similar qualitative results have been gotten for an isotype 2D lateral pP heterojunction (see SD).

\section{Conclusions}
In summary, we have developed a comprehensive technique for analyzing the electrostatics of 2D lateral homojunctions and heterojunctions, both isotype or anisotype, based on the solution of the non-linear Poisson equation in a conformal space. The model provides a suitable tool to investigate the effective depletion width, in-plane and out-of-plane electrostatic potential, electric field, and surface charge carrier distribution in dependence on the chemical doping densities, effective dielectric constant of the surrounding environment and temperature. The proposed technique could be helpful for 2D lateral junctions design and as a benchmarking for compact model development. 

\section{Acknowledgments}
This project has received funding from the European Union's Horizon 2020 research and innovation programme under grant agreements No GrapheneCore1 696656 and No GrapheneCore2 785219, the Department d'Universitats and the Ministerio de Econom\'ia y Competitividad of Spain under grant TEC2015-67462-C2-1-R (MINECO/FEDER).

\end{document}


\title{Supplementary Data: Electrostatics of Two-Dimensional Lateral Junctions}
\author{Ferney Chaves and  David Jim\'enez}
\affil {Departament d\textquoteright{}Enginyeria Electr\`onica, Escola d\textquoteright{}Enginyeria, Universitat Aut\`onoma de Barcelona, Campus UAB, 08193 Bellaterra, Spain.}
\renewcommand\Affilfont{\itshape\small}

\maketitle
\makeatletter 
\renewcommand{\thefigure}{S\@arabic\c@figure}
\renewcommand{\theequation}{S\@arabic\c@equation}
\makeatother

\section{Definition of the depletion width}

\begin{figure}[h!]
\centering
\includegraphics[scale=0.4]{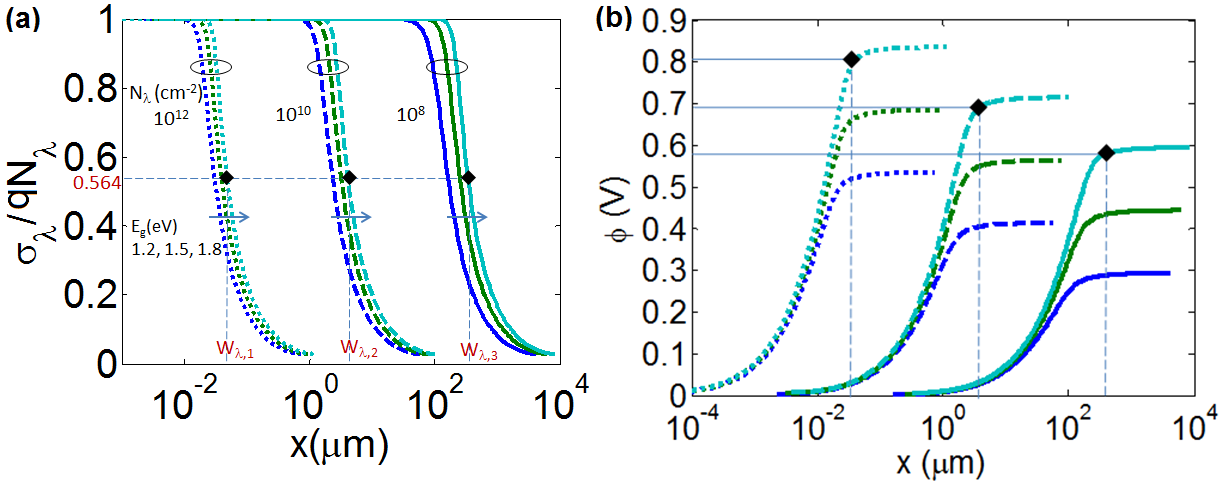}
\caption{Electrostatics of symmetric 2D $pn$ homojunctions with different band gap energies and several doping densities. (a) Surface charge density relative to the fixed charge at both sides of the junction. (b) In-plane electrostatic potential $\phi(x)$. Black marks indicate the effective depletion widths for the case $E_g=1.8$ eV.}
\label{S1}
\end{figure}

\begin{figure}[h!]
\centering
\includegraphics[scale=0.4]{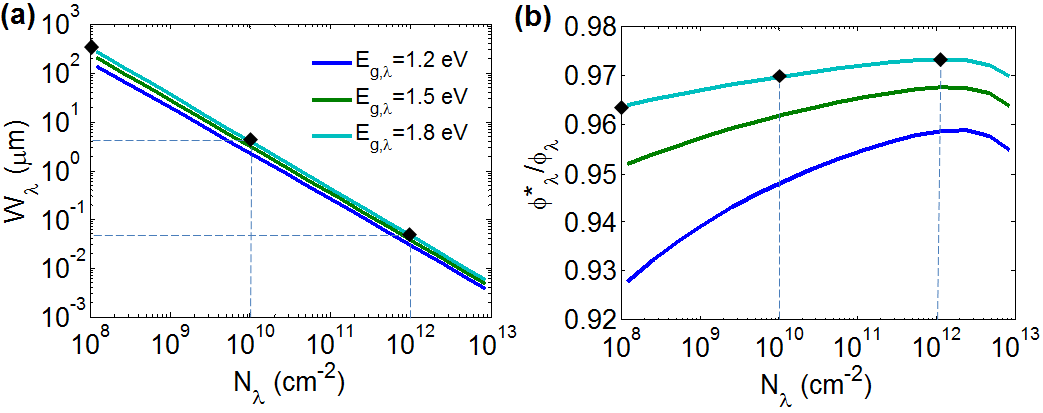}
\caption{Electrostatics of symmetric 2D $pn$ homojunctions with different band gap energies, as a function of the doping density. (a) Depletion region width. (b) Electrostatic potential at the depletion edge, relative to the asymptotic potential. Black marks indicate the effective depletion widths and relative potential for the case $E_g=1.8$ eV. }
\label{S2}
\end{figure}
\clearpage

\section{Behavior of the ratio $\gamma_\lambda=W_\lambda'/W_\lambda$}

\begin{figure}[h!]
\centering
\includegraphics[scale=0.4]{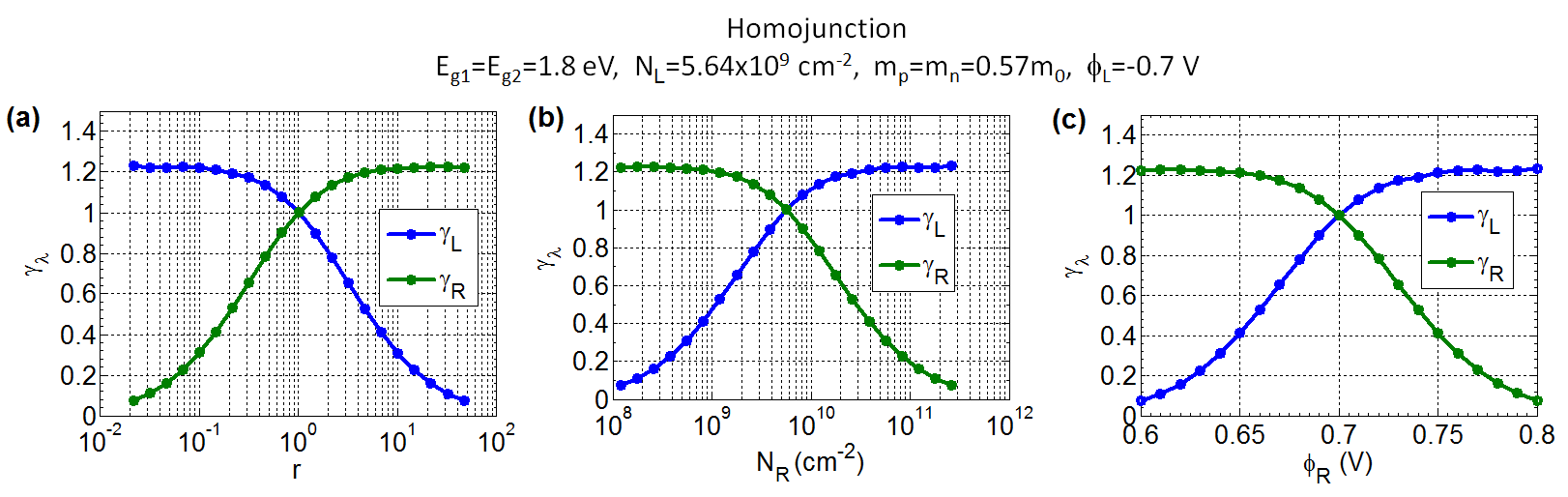}
\caption{ Ratio between the  effective depletion width extracted in this work ($W_\lambda'$) and the analytical prediction for the symmetric case ($W_\lambda$) in a $pn$ homojunction with equal electron and hole effective masses, as a  function of the asymmetry factor (a); doping density (b); asymptotic potential (c).}
\label{S5}
\end{figure}

\begin{figure}[h!]
\centering
\includegraphics[scale=0.4]{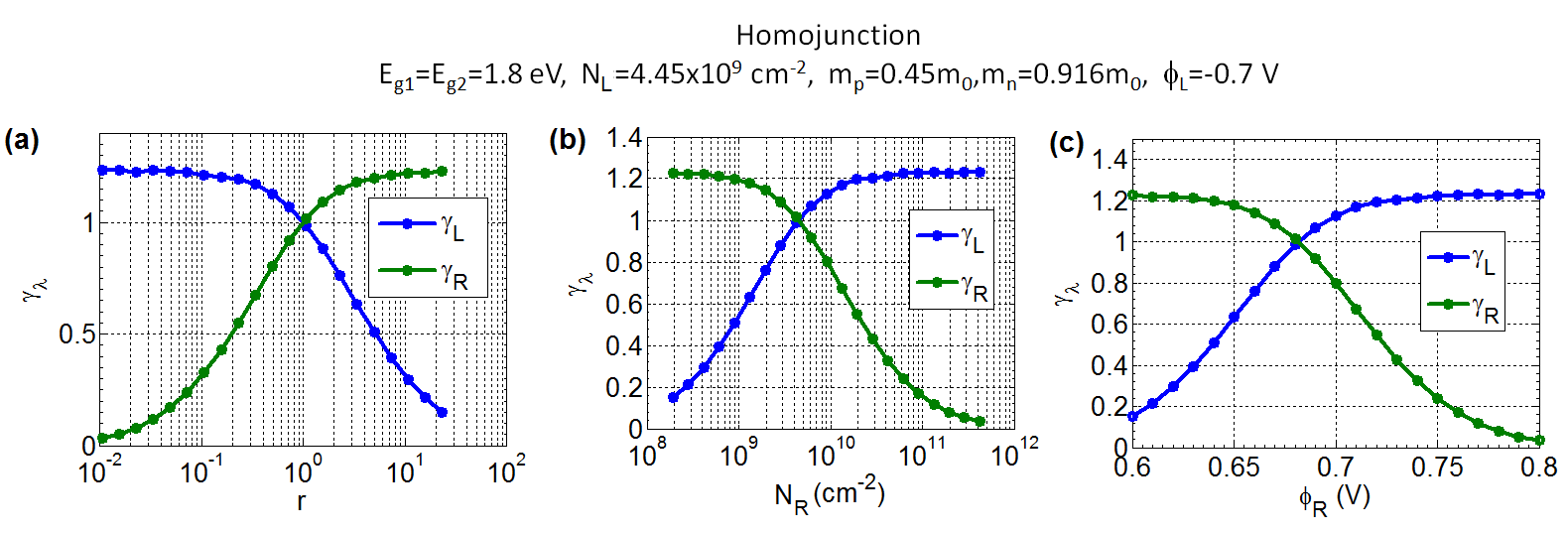}
\caption{   Ratio between the effective depletion width extracted in this work ($W_\lambda'$) and the analytical prediction for the symmetric case ($W_\lambda$) in a $pn$ homojunction with different electron and hole effective masses, as a  function of the asymmetry factor (a); doping density (b); asymptotic potential (c).}
\label{S6}
\end{figure}

\begin{figure}[h!]
\centering
\includegraphics[scale=0.4]{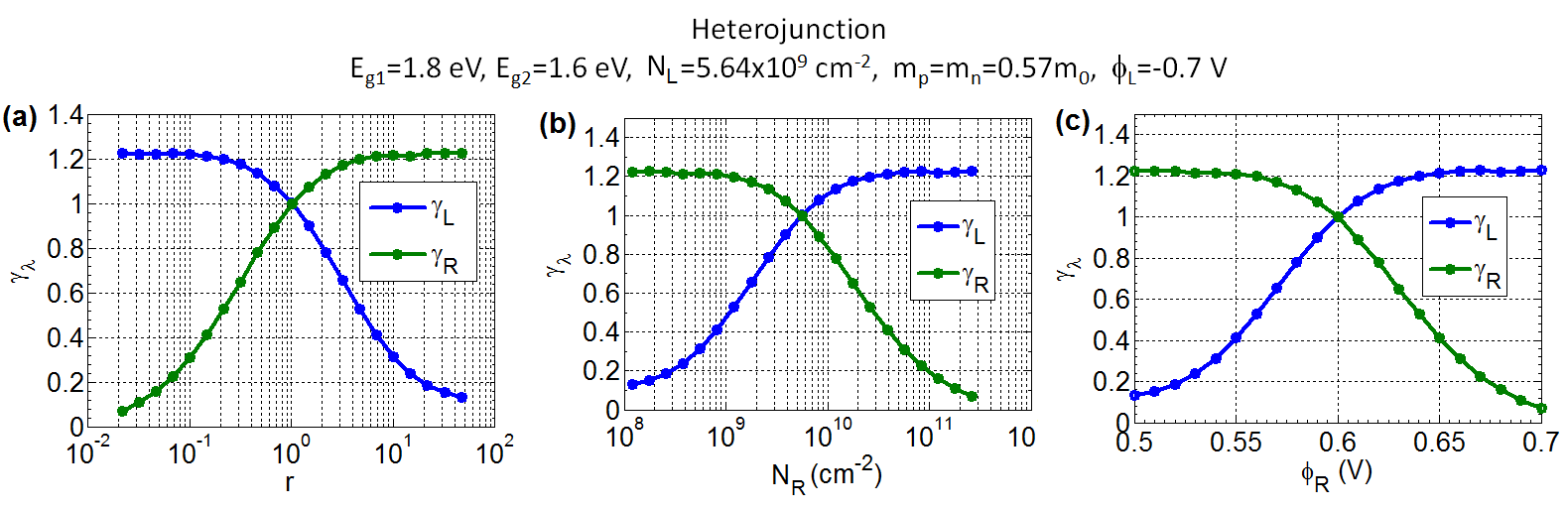}
\caption{  Ratio between the effective depletion width extracted in this work ($W_\lambda'$) and the analytical prediction for the symmetric case ($W_\lambda$) in a $pn$ heterojunction with equal electron and hole effective masses, as a  function of the asymmetry factor (a); doping density (b); asymptotic potential (c).}
\label{S7}
\end{figure}

\begin{figure}
\centering
\includegraphics[scale=0.4]{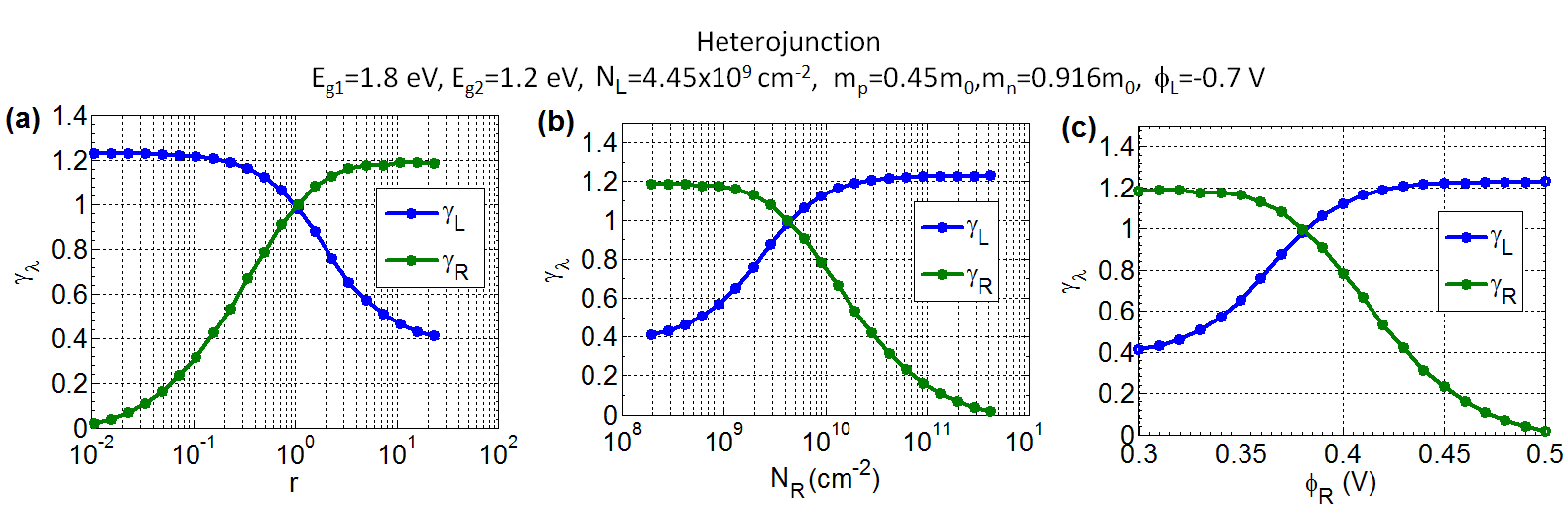}
\caption{  Ratio between the effective depletion width extracted in this work ($W_\lambda'$) and the analytical prediction for the symmetric case ($W_\lambda$) in a $pn$ heterojunction with different electron and hole effective masses, as a  function of the asymmetry factor (a); doping density (b); asymptotic potential (c).}
\label{S8}
\end{figure}

\begin{figure}[h!]
\centering
\includegraphics[scale=0.4]{FigureS7}
\caption{Dependence of $\gamma_\lambda$ with the temperature of a homojunction and a heterojunction. For the homojunction we have assumed $r$=0.15, $E_g=1.8$ eV and $m_p=m_n=0.57m_0$. For the heterojunction we have assumed $r$=0.44, $E_{g,L}=1.8$ eV, $E_{g,R}=1.2$ eV, $m_p=0.45m_0$ and $m_n=0.916m_0$}
\label{Ss5}
\end{figure}

\clearpage
\section{Electrostatics of the Heterojunction}

\begin{figure}[h!]
\centering
\includegraphics[scale=0.4]{FigureS8}
\caption{ Electrostatics of an isotype 2D lateral $pP$ heterojunction with equal electron and hole effective masses. (a) Profile of the bands. (b) In-plane electrostatic potential. (c) Equipotential lines and electric field in the $xz$-plane. (d) Total charge density relative to the fixed charge.}
\label{S9}
\end{figure}